\documentclass[journal]{IEEEtran}

\newcommand{\avec}{{\bf{a}}}

\newcommand{\evec}{{\bf{e}}}

\newcommand{\pvec}{{\bf{p}}}

\newcommand{\uvec}{{\bf{u}}}
\newcommand{\wvec}{{\bf{w}}}

\newcommand{\zerovec}{{\bf{0}}}

\newcommand{\alphavec}{{\bf{\alpha}}}

\newcommand{\Lambdamat}{{\bf{\Lambda}}}

\newcommand{\Gammamat}{{\bf{\Gamma}}}
\newcommand{\Amat}{{\bf{A}}}
\newcommand{\Bmat}{{\bf{B}}}
\newcommand{\Cmat}{{\bf{C}}}
\newcommand{\Dmat}{{\bf{D}}}
\newcommand{\Emat}{{\bf{E}}}

\newcommand{\Jmat}{{\bf{J}}}
\newcommand{\Imat}{{\bf{I}}}
\newcommand{\Lmat}{{\bf{L}}}

\newcommand{\Qmat}{{\bf{Q}}}

\newcommand{\Smat}{{\bf{S}}}

\newcommand{\Umat}{{\bf{U}}}

\newcommand{\Wmat}{{\bf{W}}}

\newcommand{\define}{\stackrel{\triangle}{=}}

\newcommand{\Psimat}{\mbox{\boldmath $\Psi$}}

\def\bsigma{{\mbox{\boldmath $\Sigma$}}}

\def\psivec{{\mbox{\boldmath $\psi$}}}

\def\alphavec{{\mbox{\boldmath $\alpha$}}}

\def\thetavec{{\mbox{\boldmath $\theta$}}}

\def\muvec{{\mbox{\boldmath $\mu$}}}

\def\muvecsmall{{\mbox{\boldmath {\scriptsize $\mu$}}}}

\def\thetavecsmall{{\mbox{\boldmath {\scriptsize $\theta$}}}}
\def\sigmatilde{\bsigma_{\mbox{\boldmath {\scriptsize $\tilde{\theta}$}}}}

\newcommand{\be}{\begin{equation}}
\newcommand{\ee}{\end{equation}}
\newcommand{\beqna}{\begin{eqnarray}}
\newcommand{\eeqna}{\end{eqnarray}}

\usepackage{amssymb, amsmath,rangecite}
\newtheorem{theorem}{Theorem}

 \usepackage[dvips]{color} 
\usepackage{lscape,longtable,nomencl}
\usepackage{comment}
\input{psfig.sty}
 \usepackage{epsfig,subfigure}
\usepackage[ruled,vlined]{algorithm2e}
\usepackage{enumitem}

\title{Power Systems  Topology and State Estimation by Graph Blind Source Separation }
\author{Sivan Grotas~\IEEEmembership{Student Member,~IEEE}, Yair~Yakoby, Idan~Gera, and Tirza~Routtenberg \IEEEmembership{Senior Member,~IEEE}
\thanks{Sivan Grotas, Yair~Yakoby, Idan~Gera, and Tirza~Routtenberg are with Department of Electrical and Computer Engineering,   Ben-Gurion University of the Negev,
 Beer-Sheva 84105, Israel.  e-mail: \{sivangr,yairyak,idange\}@post.bgu.ac.il,~tirzar@bgu.ac.il.
This research was partially supported by the 
ISRAEL SCIENCE FOUNDATION (ISF), Grant No. 1173/16
and by the BGU Cyber Security Research Center. 
The work of Sivan Grotas  was supported under a grant from 
the Ministry of Science and Technology of Israel.
 }
}

\begin{document}
	\maketitle
 \begin{abstract}
In this paper, we consider the problem of blind estimation of states and topology (BEST) in power systems. We use the linearized DC model of real power measurements with unknown voltage phases (i.e. states) and an unknown admittance matrix (i.e. topology) and show that the BEST problem can be formulated as a blind source separation (BSS) problem with a weighted Laplacian mixing matrix.  We develop the constrained maximum likelihood (ML) estimator of the Laplacian matrix for this graph BSS (GBSS) problem with Gaussian-distributed states. The ML-BEST is shown to be only a function of the states' second-order statistics. Since the topology recovery stage of the ML-BEST approach results in a high-complexity optimization problem, we propose two low-complexity methods to implement it: (1)  Two-phase topology recovery, which is based on solving the relaxed convex optimization and then finding the closest Laplacian matrix, and (2) Augmented Lagrangian topology recovery. 
We derive a closed-form expression for the associated Cram$\acute{\text{e}}$r-Rao bound (CRB) on the topology matrix estimation. 
The performance of the proposed methods is evaluated for  the IEEE-14 bus test-case system and  for a random network. It is shown that, asymptotically, the state estimation performance of the proposed ML-BEST methods coincides with the oracle’s minimum mean-squared-error (MSE) state estimator, and the MSE of the topology estimation  achieves the proposed CRB.
\end{abstract}

\section{Introduction}
State estimation is a critical component of modern energy management systems (EMSs) for multiple monitoring purposes, including analysis, security, control, situational awareness, stability assessment, power market design, and optimization of electricity dispatchment \cite{Abor,Giannakis_Wollenberg_2013}. In the DC model, the states are the bus voltage angles, while the grid topology includes the arrangement of loads or generators, transmission lines, transformers, and the statuses of system devices. 
It should be noted that this definition generalizes the computer science graph theory definition, which refers to the connectivity of the graph, since here the topology also includes the weights.
 In currently applied systems, it is assumed that the EMS has precise knowledge of the grid topology \cite{Abor}, which is used for obtaining accurate state estimation. However, knowledge of grid topology may not be available and it may change over time due to failure, opening and closing of switches on power lines, and the presence of new loads and generators.
For example, large-scale penetration of distributed generation results in regular topology changes due to ad-hoc connection of many plug-and-play components. Even worse, a distribution system operator usually lacks topology information, as many of the distributed energy resources do not belong to the utility \cite{Liao_Rajagopal_2015,2018arXiv180304506C}. The topology data may also be incorrect due to malicious attacks  \cite{Kim_Tong_2013,Poor_Tajer_2012,Zussman_2016}. Thus, methods for state estimation that are not based on a known topology are crucial for obtaining a reliable system model and high power quality. 
An additional use for topology identification is event detection, such as identifying faults, line outages, and system imbalances \cite{Routtenberg_Eldar_2018,Routtenberg_Xie_Willett_Tong}. Moreover, it can be used  to secure the system from potential cyberattacks on the topology information and
to identify the potential vulnerabilities of a power system.

Several approaches to topology identification have been proposed in the literature. Detecting topological changes has been studied in {\cite{Emami_Abur_2010,Sharon_2012}} and the conditions for the detectability of topology errors are studied in \cite{Clements_Davis_1988}. Recently, a few papers have addressed blind estimation of the grid topology by observing multiple power injection supervisory control and data acquisition (SCADA) measurements {\cite{Li_Poor_Scaglione_2013,Anwar_Mahmood_Pickering_2016}}, voltage and power data obtained by phasor measurement units (PMUs) \cite{7862849,Ye_Ardakanian_Low_Yuan_2016}, voltage measurements and their associated correlations  \cite{Bolognani_Schenato_2013}, and electricity price based market data  \cite{Kekatos_Giannakis_Baldick2016}.
 In \cite{Xie_He2017}, an unobservable attack is designed based on incomplete knowledge of the system matrix, which is learned via a blind identification approach. 
 The methods proposed in \cite{Li_Poor_Scaglione_2013,Anwar_Mahmood_Pickering_2016,Kekatos_Giannakis_Baldick2016,Xie_He2017} can reveal part of the grid topology,  such as the  grid connectivity  and the eigenvectors of the topology matrix, but  they cannot reconstruct the full topology matrix with exact scaling and true eigenvalues. 
 Thus, incorporating blind source separation (BSS) techniques with the specific characteristics of a graph seems promising.

 BSS methods aim at restoring a set of unknown source signals  from a set of observed linear mixtures of these source signals  (see, e.g. 
{\cite{Comon_1994,Pham_Garat_1997,Belouchrani_Cardoso_Moulines_1997,Cardoso_1998,Tichavsky_Oja_2006,Doron_Yeredor_Tichavsky_2007,Yeredor_2010,Lahat_Cardoso_Messer_2012,Routtenberg_Tabrikian_BSS,Routtenberg_Tabrikian_BSS_FA}}),  without prior knowledge of the sources and the mixing system. The problem of BSS has been extensively investigated in the literature in the recent two decades. 
Prior works on maximum likelihood (ML) separation in BSS deal with general stationary sources  \cite{Pham_Garat_1997,Degerine_Zaidi_2004},  autoregressive (AR) sources, and  AR Gaussian mixture model  distributed sources 
\cite{Routtenberg_Tabrikian_BSS,Routtenberg_Tabrikian_BSS_FA}. The ML BSS for nonstationary structures with varying variance-profiles
was considered in \cite{Pham_Cardoso_2001}.
However,  classical BSS  solutions are ambiguous in the sense that the order, signs and scales of the original signals cannot be retrieved. These ambiguities cannot be tolerated in the considered power system problem.
In addition, usually the distributions of the states are assumed to be Gaussian due to the central limit theorem,
 while most BSS methods cannot handle Gaussian sources. 
 Therefore, new methods for BSS are required for the semiblind scenario of a Laplacian mixing matrix with  Gaussian sources,
without permutation and scaling problems.

In addition to state estimation in power systems,
the recent field of graph signal processing (GSP) \cite{8347162} has many applications, 
{\cite{Smola2003,Narang_Ortega2012,Newman_2010}}.
A major challenge in GSP is learning the graph structure from data under
Laplacian matrix constraints
(see, e.g. \cite{Pavez_Ortega_2016,7979524,7952928})
and blind deconvolution of signals on graphs \cite{2018arXiv180103862S}, aim to jointly identify the filter coefficients and the input signal. 
In future work the recent approach has the potential to be extended to general GSP applications.

In this paper, we consider the problem of state estimation and topology identification in power systems based on active power measurements. First, we show that this problem is equivalent to the problem of BSS with a weighted Laplacian mixing matrix, where the weights are determined by the branch susceptances. Then, we derive the ML blind estimation of states and topology (ML-BEST) method
 for Gaussian-distributed states,
that  incorporates the constraints of a Laplacian mixing matrix and is shown to be a second-order statistics (SOS) method.
Since the topology recovery stage of the ML-BEST estimator is shown to be  a  NP-hard
optimization problem,  we suggest two practical low-complexity methods to implement this stage:
(1) Two-Stage topology recovery, which is based on solving the relaxed convex optimization and then finding the closest Laplacian matrix, and (2)  Augmented Lagrangian topology recovery.
 Preliminary results can be found in \cite{Gera_Yakoby_GlobalSip}. We also derive a closed-form expression for the Cram$\acute{\text{e}}$r-Rao bound (CRB). Finally, simulations demonstrate that the proposed ML-BEST methods are applicable for different network topologies, and asymptotically achieve the CRB.

The remainder of the paper is organized as follows. 
In Section \ref{problem_formulation_sec} we introduce the system model
and the graph BSS (GBSS) problem
for state and topology estimation in power systems. 
The ML-BEST solution is defined and two different practical methods for its topology recovery stage are suggested in Sections \ref{ML_BEST_section} and \ref{prac}, respectively. 
Section \ref{remark_section} offers some remarks, including
 a parameter
identifiability analysis, complexity discussion, and possible extensions of the proposed model and methods.  A closed-form expression for the CRB of the topology matrix is derived in Section \ref{CRB_section}. The proposed methods are evaluated via simulations in Section \ref{simulations_section}. Conclusions appear in Section \ref{diss}.

\section{Problem formulation}
\label{problem_formulation_sec}
In this section, we formulate the problem of estimating the state and topology/admittance matrix in power systems under the 
 linear DC power model.
We show that this problem is equivalent to  BSS with a Laplacian mixing matrix.

\subsection{Notation}
In the rest of this paper vectors  are denoted by boldface lowercase letters and matrices  by boldface uppercase letters.
The $K \times K$ identity matrix is denoted
by $\Imat_K$,
and  ${\mathbf{1}}_K$  denotes the constant $K$-length one   vector.
The vectors $\zerovec$ and $\evec_m$   are zero vectors (with appropriate dimension),
 except for the 
	$m$th element of $\evec_m$, which is $1$. 
Additionally,  
	$\delta_{m,k}$  denotes Kronecker’s delta, which equals $1$ if $m=k$
and $0$ otherwise.
The notations 
 $|\cdot|$, ${\text{Tr}}\{\cdot\}$,  and $\otimes$ denote the determinant operator,
the trace operator,
and the Kronecker product, respectively.
For a full-rank matrix $\Amat$, $\Amat^\dagger\define \left(\Amat^T \Amat\right)^{-1}\Amat^T$
is the 
 Moore-Penrose pseudo-inverse.
The $m$th element of the vector $\avec$,
the $(m,q)$th element of the matrix $\Amat$,
and the $(m_1:m_2\times q_1:q_2)$ submatrix of $\Amat$ 
 are denoted by $a_m$, $\Amat_{m,q}$, and 
 $\Amat_{m_1:m_2,q_1:q_2}$, respectively.
If $\Amat$ is a positive semidefinite matrix we denote it by $\Amat\succeq \zerovec$ and its  square root, $\Amat^{\frac{1}{2}}$, satisfies
$\Amat^{\frac{1}{2}}\Amat^{\frac{1}{2}}=\Amat$,
where $\Amat^{-\frac{1}{2}}$ denotes the inverse of this square root.
For any matrix $\Amat$, 
$||\Amat||_F$ and $||\Amat||_0$ denote its Frobenius  and
 $\ell_0$-(pseudo)norm (counting
its non-zero entries),  respectively, 
 $\{\Amat\}^+=\max\{\Amat,0\}$ is the nonnegative part of  $\Amat$, 
and $ {\text{vec}} (\Amat)$
is a vector obtained by stacking its columns.
Similarly, for any symmetric matrix $\Smat$, ${\text{vech}} (\Smat)$
is a vector obtained by stacking the columns of the lower
triangular part of $\Smat$, including the diagonal, into a single column.
Finally, we denote the cone of real symmetric matrices of size
$M\times M$ by ${\mathbb{S}}^M$.
\subsection{Graph representation of power systems}
A power system can be represented as
 an undirected connected weighted graph,
${\mathcal{G}}({\mathcal{V}},\xi)$,
where the set of vertices, ${\mathcal{V}}=\{1,\ldots,M\}$,
 is the set of buses
(that represent interconnections, generators or loads)
and  the edge set, $\xi$, is the set of connected transmission lines  between the buses.
An arbitrary orientation is assigned to each edge $e_i=(m,k)\in \xi$, $m,k=1,\ldots,M$,
$k<m$, $i=1,\ldots, \frac{M(M-1)}{2}$, that are ordered in a lexicographical order,
which connects the vertices $m$ and $k$.
The cardinality of the edge set, $|{\xi}|=\frac{M(M-1)}{2}$, represents all possible connections in the graph.
According to  the $\pi$-model of transmission lines \cite{Abor},
each line is characterized by the line admittance $Y_{m,k}$, $\forall (m,k) \in \xi$.

The incidence matrix of a graph is
$\Bmat\in{\mathbb{R}}^{M\times \frac{M(M-1)}{2}}$ \cite{Newman_2010},
where the $(m,i)$ element of $\Bmat$ is given by
\beqna
   \label{Bmat_definition}
\Bmat_{m,i} = 
     \left\{\begin{array}{ll}
       1 & e_i = (m, k){\text{ is connected, $m$ is the source}} \\
			 -1 & e_i = (k, m){\text{ is connected, $k$ is the source}} \\
       0 & {\text{otherwise}}\\
     \end{array}\right.,
\eeqna
$\forall m=1,\ldots,M$ and $i=1,\ldots, \frac{M(M-1)}{2}$.
In addition,
let $\Gammamat  \in {\frac{M(M-1)}{2}\times \frac{M(M-1)}{2}}$ be a diagonal matrix where
 $\Gammamat_{i,i} = Y_{m,k}$ if
$e_i = (m,k)$, $i=1,\ldots, \frac{M(M-1)}{2}$.
 For connections that do not exist we use $\Gammamat_{i,i} =0$.
Then, we can define the graph Laplacian matrix, $\Lmat$,  as
\be
\label{Lap_matrix}
\Lmat \define \Bmat\Gammamat\Bmat^T.
\ee
The matrix $\Lmat\in{\mathbb{R}}^{M\times M}$
is a real, symmetric, and positive semidefinite matrix\footnote{It should be noted that $\Lmat$ is a positive semidefinite matrix, assuming we only have positive susceptances \cite{Gustavsen_Semlyen2001}.}, which satisfies
the null space property, $\Lmat\mathbf{1}_M=\zerovec$,
and with nonpositive off-diagonal elements.

\subsection{DC model and problem formulation}
\label{BSS_original}
We consider the DC power flow model \cite{Abor}, which is based on the following assumptions on the network:
\begin{description}
\item[A.1] Branches are considered lossless, which results in 
$Y_{m,k}=b_{m,k}$,
where $b_{m,k}$ is the susceptance of the $(m,k)$
branch.
\item[A.2] The bus voltage magnitudes, $V_m$, $m=1,\ldots,M$, are approximated by 1 per unit (p.u.).
\item[A.3] Voltage angle differences across branches are small, such that $\sin(\theta_m-\theta_k) \approx \theta_m-\theta_k$, where  $\theta_m$, $m=1,\ldots,M$, are the bus voltage angles. 
\end{description}
Under Assumptions A.1-A.3,
the active power injected at bus $m$ satisfies 
\beqna
\label{power_bus}
p_m
=-\sum\nolimits_{k=1}^M b_{m,k} V_m V_k \sin(\theta_m-\theta_k)\hspace{1.4cm}
\nonumber\\
\approx -\sum\nolimits_{k=1}^M Y_{m,k}(\theta_m-\theta_k),~\forall m=1,\ldots,M.
\eeqna

Now, let $\pvec[n]\define{[p_1[n],\dots,p_M[n]]}^T$
be the vector of active power injected 
and $\thetavec[n]\define{[\theta_1[n],\dots,\theta_M[n]]}^T$  the vector of voltage phase angles at time $n$, $\forall n=0,\ldots,N-1$.
Then, based on the model from (\ref{power_bus}),
the noisy linearized DC model of the network can be written as
\be
\label{the_model_0}
\pvec[n] =\Lmat\thetavec[n]+\wvec[n],~~~n=0,\ldots,N-1,
\ee
where the topology  matrix $\Lmat$,   defined in (\ref{Lap_matrix}),
  is a deterministic unknown Laplacian  matrix, which
is considered
 static for a short-period of time and under normal operating conditions.
The noise  is a stationary Gaussian sequence  with zero mean  and a covariance matrix $\sigma^2\Imat_M$, i.e.  $\wvec [n]\sim{\cal{N}}(\zerovec,\sigma^2\Imat_M)$, and it is assumed that the additive noises  are independent of the state vectors. 
The vectors $\{\thetavec[n]\}$, $n=0,\ldots,N-1$, are assumed to be unknown  random  states
with a joint probability density function (pdf) $f_\thetavecsmall(\cdot)$
and  marginal pdfs  of $\theta_m$, $f_{\theta_m}(\cdot)$, $m=1,\ldots,M$.
By subtracting the mean from the data, we can assume, without loss of generality, that  $\thetavec$ has zero mean. The resulting centralized measurements are given by ${\pvec}[n]-\bar{\pvec}$,
where $\bar{\pvec}\define \frac{1}{N}\sum_{n=0}^{N-1} {\pvec}[n]$ is the sample mean.
For the rest of this paper, ${\pvec}[n]$ will denote the mean-centered active power data.

Now, in order to reformulate the model with a full-rank mixing matrix, we use the relation
\be
\label{relation}
\Lmat=\Umat \tilde{\Lmat} \Umat^T,
\ee
where
\be
\label{AAA}
\Umat\define\left[\begin{array}{c}-{\mathbf{1}}_{M-1}^T\\
\Imat_{M-1}\end{array}\right]\in{\mathbb{R}}^{M\times(M-1)}
\ee
and
 $\tilde{\Lmat}\define \Lmat_{2:M,2:M}$ is a $1$st-order reduced-Laplacian matrix,
which is obtained by  removing
the first row and first column of $\Lmat$.
By substituting (\ref{relation}) in (\ref{the_model_0}), one obtains
\be
\label{General_graph3}
\pvec[n] =\Umat\tilde{\Lmat}\tilde{\thetavec}[n]+\wvec[n],~~~n=0,\ldots,N-1,
\ee
where
\beqna
\tilde{\thetavec}[n]\define \Umat^T{\thetavec}[n]
=[\theta_2[n]-\theta_1[n],\ldots,\theta_M[n]-\theta_1[n]],\nonumber
\eeqna
$n=0,\ldots,N-1$.
By multiplying both sides of (\ref{General_graph3}) with $ \Umat^\dagger$, it can be verified that the model in
(\ref{General_graph3}) is equivalent to
\be
\label{General_graph4}
\tilde{\pvec}[n] =\tilde{\Lmat}\tilde{\thetavec}[n]+\tilde{\wvec}[n],~~~n=0,\ldots,N-1,
\ee
where $\tilde{\pvec}[n]\define \Umat^\dagger \pvec[n]$ and $\tilde{\wvec}[n]\define \Umat^\dagger \wvec[n]$, $n=0,\ldots,N-1$.
In addition,
it can be shown (see, e.g. pp. 134-144 \cite{Papoulis}) that the modified noise sequence satisfies 
$\tilde{\wvec} [n]\sim{\cal{N}}(\zerovec,\sigma^2\Umat^\dagger (\Umat^\dagger)^T )$, $n=0,\ldots,N-1$.

We assume here that  all sources are time-independent Gaussian distributed, i.e.
$\thetavec [n]\sim{\cal{N}}(\zerovec,\bsigma_\thetavecsmall)$, $n=0,\ldots,N-1$.
Thus, 
 ${\tilde{\thetavec}}[n]\sim{\cal{N}}(\zerovec,\sigmatilde)$, $n=0,\ldots,N-1$,
where 
$
\sigmatilde\define \Umat^T\bsigma_\thetavecsmall \Umat
$.
Under the assumption  that $\bsigma_\thetavecsmall$ is known,
the observations
vectors are also independent Gaussian-distributed vectors,
i.e.  ${\pvec}[n]\sim{\cal{N}}\left(\zerovec,\bsigma_\pvec({\Lmat},\sigma^2)\right)$
and $\tilde{\pvec}[n]\sim{\cal{N}}\left(\zerovec,\bsigma_{\tilde{\pvec}}(\tilde{\Lmat},\sigma^2)\right)$, $n=0,\ldots,N-1$, 
where
\be
\label{cov_p}
\bsigma_\pvec({\Lmat},\sigma^2)
 \define{\Lmat}^{T}\bsigma_\thetavecsmall{\Lmat}+\sigma^2\Imat_M
\ee
and, assuming nonsingular matrices,
\be
\label{cov_tilde}
\bsigma_{\tilde{\pvec}}(\tilde{\Lmat},\sigma^2)
 \define\tilde{\Lmat}^{T}\sigmatilde \tilde{\Lmat}
+{\sigma}^2 \Umat^\dagger  (\Umat^\dagger)^T
.
\ee

The  reduced topology matrix, $\tilde{\Lmat}$, has the following properties \cite{Newman_2010,7979524}:
\begin{description}
  \item[P.1] Full rank - Under the assumption of a connected graph,  $\tilde{\Lmat}$ is a nonsingular matrix of rank $M-1$ and, thus, can be identified in general.
	In power system terminology, we assume that there are no unobservable islands in the grid.
  \item[P.2] Positive semidefinite - Since $\Lmat$ is a symmetric, positive semidefinte  matrix,  $\tilde{\Lmat}$ is also a symmetric, positive semidefinte matrix.
	 \item[P.3] Nonpositive off-diagonal elements - $\tilde{\Lmat}_{k,m}\leq 0$, $\forall k,m=1,\ldots,M-1$, $k\neq m$.
  \item[P.4] Diagonally dominant - Since $\Lmat$ is a Laplacian matrix, 
$\tilde{\Lmat}$ is a diagonally dominant matrix, i.e. 
$\sum_{m=1,m\neq k}^{M-1}|\tilde{\Lmat}_{k,m}|\leq |\tilde{\Lmat}_{k,k}|$, $\forall k=1,\ldots,M-1$.
 \item[P.5] Sparsity (optional) -  It is shown in previous works
  that the power system is sparse  \cite{Wang_Scaglione_Thomas_2010}, i.e.
	the zero pseudonorm of the off-diagonal entries of $\tilde{\Lmat}$,
$||\tilde{\Lmat}||_{0-{\text{off}}}$, is much smaller than $(M-1)(M-2)$. 
\end{description}

\section{ML-BEST}
\label{ML_BEST_section}
In this section, we develop the basic ML-BEST approach that jointly reconstructs the matrix $\Lmat$ and the states $\thetavec[n]$, $n=0,\ldots,N-1$, for the model from Section \ref{problem_formulation_sec}. This problem can be interpreted as a BSS problem with a Laplacian mixing matrix, or graph BSS (GBSS).
 First, in Subsection \ref{MMSE_subsection} the minimum mean-squared-error (MMSE) estimator of the random states, ${\thetavec}[n]$, $n=0,\ldots,N-1$, is developed. Then, in Subsection  \ref{mixing_subsection}, we develop the ML estimator of the noise variance, $\sigma^2$, and formulate the optimization problem describing the ML estimator of the mixing system.

	\subsection{MMSE state estimation}
	\label{MMSE_subsection}
For given $\Lmat$ and  $\sigma^2$, the sequences
$\pvec[n]$, $n=0,\ldots,N-1$, $\thetavec[n]$, $n=0,\ldots,N-1$, 	are
jointly Gaussian. Thus, in this case the MMSE
estimator of the state vector  is a linear estimator given by
	(see, e.g. Chapter 20 in \cite{Matrix_book}, \cite{Kosut_Jia_Thomas_Tong_2011})
\beqna
\label{est_singular}
\hat{\thetavec}[n]= \bsigma_\thetavecsmall \Lmat 
\left({\Lmat}^{T}\bsigma_\thetavecsmall{\Lmat}+\sigma^2\Imat_M\right)^{\dagger}
\pvec[n],
\eeqna
	$n=0,\ldots,N-1$. 
We refer to the estimator in \eqref{est_singular} as the 
	oracle MMSE state estimator,
	 i.e. an ideal estimator which has perfect knowledge of the noise variance and the  system topology.
	
	The practical state estimator for the considered GBSS problem is obtained by
plugging in the ML estimators of the noise variance and the
 reduced-Laplacian matrix, $\hat{\sigma}^2$ and  $\hat{\tilde{\Lmat}}^{\text{ML}}$, respectively,
that are developed in the following in Subsections \ref{noise_sub} and \ref{mixing_subsection}, into (\ref{est_singular}), which results in
\beqna
\label{est_singular5}
\hat{{\thetavec}}[n]=\bsigma_\thetavecsmall \hat{{\Lmat}}^{\text{ML}}
\left(\left(\hat{{\Lmat}}^{\text{ML}}\right)^T\bsigma_\thetavecsmall \hat{{\Lmat}}^{\text{ML}}+\hat{\sigma}^2 \Imat_M\right)^{\dagger}
\pvec[n],
\eeqna
	$n=0,\ldots,N-1$.

For high signal-to-noise ratio (SNR) values, i.e. when $\sigma^2\rightarrow 0$,  the matrix ${\Lmat}^{T}\bsigma_\thetavecsmall{\Lmat}$ is a singular matrix and, thus,  the  covariance matrix of the data from \eqref{cov_p} is also a singular matrix.
In this case, instead of using pseudo inverse as in \eqref{est_singular} and 
	\eqref{est_singular5},
 the unknown parameters can also be treated by removing the linearly dependent random variable (see, e.g. Chapters 3 and 10 in \cite{gallager2013stochastic}). 
In power system state estimation this is usually done by setting  one bus as a reference bus 
and setting its angle  to zero  (see, e.g. \cite{Abor}), and then only estimating ${\tilde{\thetavec}}[n]$.
	Here we prefer to use instead the state estimation method in \eqref{est_singular} and 
	\eqref{est_singular5} for estimation of $\thetavec[n]$.


\subsection{ML estimation of the noise variance}
\label{noise_sub}
It is shown in \cite{anderson1956statistical,Tong_Huang_1991,Stoica_Jansson_2009}
that for 
Gaussian measurements with the aforementioned structure, 
the ML estimator of the noise variance $\sigma^2$ is given by
\be
\label{MLvariance}
\hat{\sigma}^2=\lambda_M,
\ee
where $\lambda_1 \geq\lambda_2\geq\ldots\geq \lambda_{M}$ are the eigenvalues of the sample covariance matrix,
\be
\label{total_cov}
\hat{\bsigma}_\pvec\define
 \frac{1}{N}\sum_{n=0}^{N-1} {\pvec}[n]{\pvec}^T[n].
\ee

\subsection{System identification: ML estimation of the mixing matrix}
\label{mixing_subsection}
By using the invariance property of the ML estimator \cite{Kayestimation} and the relation in \eqref{relation},
the ML estimator
of the full Laplacian matrix can be obtained from  the ML estimator of the
 reduced-Laplacian matrix, $\hat{\tilde{\Lmat}}^{({\text{ML}})}$,
as follows:
\beqna
\label{relation_ML}
\hat{\Lmat}^{({\text{ML}})}=\Umat \hat{\tilde{\Lmat}}^{({\text{ML}})} \Umat^T.
\eeqna

In the  following,
 the ML estimation of the reduced topology matrix, $\tilde{\Lmat}$, is formulated and is shown to be NP-hard. Practical methods to approximate
 the ML estimator of $\tilde{\Lmat}$, $\hat{\tilde{\Lmat}}^{({\text{ML}})}$,
are developed in the next section.
Under the model from (\ref{General_graph4}) and the Gaussian-distributed sources assumptions, 
the  normalized log likelihood  of $\tilde{\pvec}[n]$, $n=0,\ldots,N-1$, after removing constant terms
and substituting the ML estimator of the noise variance  from \eqref{MLvariance},
satisfies
\beqna
\label{Gaussian22}
\psi(\tilde{\Lmat})
=- {\text{Tr}}
\left\{\hat{\bsigma}_{\tilde{\pvec}} \bsigma_{\tilde{\pvec}}^{-1}(\tilde{\Lmat},\hat{\sigma}^2) \right\}
-\log \left|\bsigma_{\tilde{\pvec}}(\tilde{\Lmat},\hat{\sigma}^2)\right|,
\eeqna
where
\be
\label{cov_est}
\hat{\bsigma}_{\tilde{\pvec}}\define \frac{1}{N}\sum_{n=0}^{N-1} \tilde{\pvec}[n]
\tilde{\pvec}^T[n]=\Umat^\dagger \hat{\bsigma}_{{\pvec}} (\Umat^\dagger)^T
\ee
 is the modified sample covariance matrix and the last equality is obtained by substituting \eqref{total_cov}.
That is, the log-likelihood from (\ref{Gaussian22}) depends on the data only through the  sample covariance matrix, $\hat{\bsigma}_{\tilde{\pvec}}$, which is the sufficient statistic for estimating $\tilde{\Lmat}$.

Since the reduced-Laplacian matrix satisfies
 Properties P.1-P.4, we are interested in 
minimizing  $-\psi(\tilde{\Lmat})$
 over the domain of symmetric matrices
and under the associated constraints as follows:
\beqna
\label{optimization5}
\begin{array}{l}
\min\limits_{\tilde{\Lmat}\in {\mathbb{S}}^{M-1}}-\psi(\tilde{\Lmat})\\
{\text{ such that }}\\
1)~\tilde{\Lmat}\succ \zerovec\\
2)~{\tilde{\Lmat}}_{m,k}\leq 0,~~~\forall m,k=1,\ldots,M-1,~k<m\\ 
3)~\sum_{k=1}^{M-1}\tilde{\Lmat}_{m,k}\geq 0,~~~\forall m=1,\ldots,M-1
\end{array}.
\eeqna
The Gaussian log-likelihood function, $\psi(\tilde{\Lmat})$, is a concave function of the inverse
covariance matrix, $\bsigma_{\tilde{\pvec}}^{-1}(\tilde{\Lmat},\hat{\sigma}^2)$.
 However, even without the sparsity constraint,  the constraints in \eqref{optimization5} cannot be rewritten as convex
constraints on $\bsigma_{\tilde{\pvec}}^{-1}(\tilde{\Lmat},\hat{\sigma}^2)$. 
Therefore, the resulting optimization  is
not a convex optimization and, in addition, a
direct Karush-Kuhn-Tucker (KKT) conditions \cite{Boyd_2004} solution of this constrained minimization is intractable.
Two low-complexity implementation methods are described in the next section.

Imposing directly the sparsity constraint in P.5 usually results in 
complex combinatorial searches, and
following advances in compressive sensing {\cite{Donoho_eladm,CS_Davenport_Eldar}},  the sparsity constraint can be  approximated by
restricting the off-diagonal $\ell_1$-norm.
We perform simulations that suggest that simple elementwise thresholding of the estimated Laplacian matrix is competitive with $\ell_1$ methods. 
Thus,  at the end of  the ML-BEST approach, we thresholded the off-diagonal elements of the estimator of the topology matrix, 
$\hat{\Lmat}^{({\text{ML}})}$,  from \eqref{relation_ML}, with a threshold, $\tau$,
 such that the $(k,m)$th  element  of the final estimation is given by
\beqna
\label{sparse_ML}
\hat{{{\Lmat}}}^{({\text{ML}})}_{k,m}=
\left\{\begin{array}{lr}
\hat{{{\Lmat}}}^{({\text{ML}})}_{k,m}&{\text{if }}|\hat{{{\Lmat}}}^{({\text{ML}})}_{k,m}|>\tau\\
0&{\text{otherwise}}\end{array}\right.,
\eeqna
$k,m=1,\ldots,M-1$, $k\neq m$.
The threshold $\tau$ should  be tuned until the desired level of sparsity is achieved, while keeping connectivity. The diagonal elements of ${\Lmat}$ are known to be positive for the Laplacian matrix, which thus, has partially known support.
Thus,  $\tau$ set to be smaller than the magnitude of the smallest estimated element of the diagonal:
\be
\label{tau}
\tau=\alpha\min_{m=1,\ldots,M}\hat{{{\Lmat}}}^{({\text{ML}})}_{m,m},
\ee
where $0<\alpha<1$.
The value of $\alpha$ can be set to the inverse of the number of buses, $\frac{1}{M}$,
or of the average nodal degree \cite{Newman_2010}.

The basic ML-BEST algorithm is summarized in Algorithm \ref{Alg0},
 for any  method of estimation of
the reduced-Laplacian matrix, ${\tilde{\Lmat}}$.
 Two such methods are described in Section \ref{prac}.


\begin{algorithm}
 \SetAlgoLined
{\bf{Input: }}\begin{itemize}
\item {Observations $\pvec[n]$, $n=0,\ldots,N-1$}.
\item State covariance matrix, $\bsigma_{\tilde{\theta}}$.
\end{itemize}
 {\bf{Output:}} Estimators $\hat{\Lmat}$ and $\hat{\thetavec}[n]$, $n=0,\ldots,N-1$.\\
{\bf{Algorithm Steps:}}
\begin{enumerate}
\item (Optional) Remove the sample mean, $\bar{\pvec}\define \frac{1}{N}\sum_{n=0}^{N-1} {\pvec}[n]$, from the observations 
$\pvec[n]$, $n=0,\ldots,N-1$.
\item Obtain the sample covariance matrix, $\hat{\bsigma}_\pvec$, by (\ref{total_cov}).
\item 
Perform eigendecomposition operation for the sample
covariance matrix
 $\hat{\bsigma}_\pvec$ to find
 its eigenvalues $\lambda_1 \geq\lambda_2\geq\ldots\geq \lambda_{M}$.
\item  Estimate the noise variance  by  the
smallest eigenvalue,
$
\hat{\sigma}^2=\lambda_M$.
\item Estimate the reduced-Laplacian matrix
 and obtain approximation to $\hat{\tilde{\Lmat}}^{({\text{ML}})}$,
 for example,  by the two-phase/augmented ML-BEST from Section \ref{prac}.
\item Reconstruct the full topology matrix according to \eqref{relation_ML}:
\[\hat{\Lmat}^{({\text{ML}})}=\Umat \hat{\tilde{\Lmat}}^{({\text{ML}})} \Umat^T.\]
\item  Impose sparsity  by 
setting the threshold according  to \eqref{tau}:
\[
\tau=\alpha
\min_{m=1,\ldots,M}\hat{{{\Lmat}}}^{({\text{ML}})}_{m,m}
\]
and thresholding such that the $(k,m)$th element of the final estimation is
given by \eqref{sparse_ML}:
\[
\hat{\tilde{{\Lmat}}}^{({\text{ML}})}_{k,m}=
\left\{\begin{array}{lr}
\hat{\tilde{{\Lmat}}}^{({\text{ML}})}_{k,m}&{\text{if }}|\hat{\tilde{{\Lmat}}}^{({\text{ML}})}_{k,m}|>\tau\\
0&{\text{otherwise}}\end{array}\right.,
\]
$k,m=1,\ldots,M-1$, $k\neq m$.
\item Evaluate the sources according to (\ref{est_singular5}):
\[
\hat{{\thetavec}}[n]=\bsigma_\thetavecsmall \hat{{\Lmat}}^{\text{ML}}
\left(\left(\hat{{\Lmat}}^{\text{ML}}\right)^T\bsigma_\thetavecsmall \hat{{\Lmat}}^{\text{ML}}+\hat{\sigma}^2 \Imat_M\right)^{\dagger}
\pvec[n],
\]
$n=0,\ldots,N-1$.
\end{enumerate}
 \caption{Basic ML-BEST Algorithm}
\label{Alg0}
\end{algorithm}

\section{Practical implementations of the  ML-BEST}
\label{prac}
In this section,  two  low-complexity estimation methods of the reduced topology are derived: 1) Two-Stage topology recovery in Subsection \ref{two_phase_CML}; and 2) Augmented Lagrangian topology recovery in Subsection \ref{augmented}.

\subsection{Two-phase topology recovery}
\label{two_phase_CML}
In this subsection, we propose a low-complexity method for solving (\ref{optimization5}) in two phases.
First, we
relax the original optimization problem from (\ref{optimization5}), by removing constraints $2)$ and $3)$  into 
\beqna
\label{optimization5_rel}
\begin{array}{l}
\min\limits_{\tilde{\Lmat}\in {\mathbb{S}}^{M-1}}-\psi(\tilde{\Lmat})
{\text{ such that }}1)~\tilde{\Lmat}\succeq \zerovec
\end{array}.
\eeqna
It is well known that the relaxed optimization problem from (\ref{optimization5_rel}) is a convex optimization w.r.t.
${\bsigma}_{\tilde{\pvec}}^{-1}(\tilde{\Lmat},\hat{\sigma}^2)$ and
 the optimal solution is the sample covariance matrix inverse,
$\hat{\bsigma}_{\tilde{\pvec}}^{-1}$,
under the assumption of nonsingular matrices (see, e.g.   p. 466 in \cite{Horn_Johnson_book}, \cite{Zhang_Wiesel_Greco_2013}).
Then, by using the invariance property of the ML estimator \cite{Kayestimation},
 the one-to-one mapping in (\ref{cov_tilde}),
and the symmetry $\tilde{\Lmat}^{T}=\tilde{\Lmat}$,  one obtains
that the unique minimum of \eqref{optimization5_rel} w.r.t. $\tilde{\Lmat}$
, which is the ML estimator of a symmetric positive definite mixing matrix, 
$\hat{\tilde{\Lmat}}^{\text{PD}}$, satisfies
\beqna
\label{Gaussian3}
\hat{\bsigma}_{\tilde{\pvec}}
=\hat{\tilde{\Lmat}}^{\text{PD}}\sigmatilde\hat{\tilde{\Lmat}}^{\text{PD}}
+\hat{\sigma}^2\Umat^\dagger  (\Umat^\dagger)^T,
\eeqna
which implies  that
\be
\label{H_estimation}
\hat{\tilde{\Lmat}}^{\text{PD}}=\sigmatilde^{-\frac{1}{2}}\left(\sigmatilde^{\frac{1}{2}}
\left(\hat{\bsigma}_{\tilde{\pvec}}-\hat{\sigma}^2\Umat^\dagger  (\Umat^\dagger)^T\right)
\sigmatilde^{\frac{1}{2}}\right)^{\frac{1}{2}}\sigmatilde^{-\frac{1}{2}}.
\ee

In the second phase, we 
find the closest  graph Laplacian   matrix  to the matrix 
$\Umat\hat{\tilde{\Lmat}}^{\text{PD}}\Umat^T$
 in the sense of Frobenius norm. 
Thus, we solve
 the following  optimization problem:
\beqna
\label{optimization_after_ML}
\begin{array}{l}
\min\limits_{{\Lmat}\in {\mathbb{S}}^{M}}||\Umat\hat{\tilde{\Lmat}}^{\text{PD}}\Umat^T-\Lmat||_F\\
{\text{ such that }}\\
1)~{\Lmat}\succeq \zerovec\\
2)~{\Lmat}_{m,k}\leq 0,~~~\forall m,k=1,\ldots,M,~k<m\\ 
3)~\sum_{k=1}^{M} {\Lmat}_{m,k}= 0,~~~\forall m=1,\ldots,M
\end{array}.
\eeqna
The problem in (\ref{optimization_after_ML})
is a convex optimization problem and
can be efficiently computed by standard
semidefinite program solvers,  such as CVX \cite{CVX}.
This two-phase topology recovery algorithm is summarized in Algorithm \ref{Alg1}.
The ML-BEST approach with two-phase topology recovery is implemented by Algorithm \ref{Alg0}, where Step 5 is implemented by Algorithm \ref{Alg1}.
\begin{algorithm}[h]
 \SetAlgoLined
{\bf{Input: }}
$\sigmatilde$, $\hat{\bsigma}_{{\pvec}}$, $\hat{\sigma}^2$.\\
 {\bf{Output:}} Estimator $\hat{\tilde{\Lmat}}^{\text{ML}}$.\\
{\bf{Algorithm Steps:}}
\begin{enumerate}
\item Evaluate  the reduced sample covariance matrix
  from (\ref{cov_est})
by $\hat{\bsigma}_{\tilde{\pvec}}=\Umat^\dagger \hat{\bsigma}_{{\pvec}} (\Umat^\dagger)^T$.
\item Evaluate the optimal solution of the optimization in \eqref{optimization5_rel} by \eqref{H_estimation}:
\[\hat{\tilde{\Lmat}}^{{\text{PD}}}=\sigmatilde^{-\frac{1}{2}}\left(\sigmatilde^{\frac{1}{2}}
\left(\hat{\bsigma}_{\tilde{\pvec}}-\hat{\sigma}^2\Umat^\dagger  (\Umat^\dagger)^T\right)
\sigmatilde^{\frac{1}{2}}\right)^{\frac{1}{2}}\sigmatilde^{-\frac{1}{2}}.\]
\item Find the closest Laplacian matrix, $\hat{{\Lmat}}^{({\text{ML}})}$,
to $\Umat\hat{\tilde{\Lmat}}^{\text{PD}}\Umat^T$,
by solving the convex optimization  problem in (\ref{optimization_after_ML})
(by
solvers such as CVX \cite{CVX}).
\end{enumerate}
 \caption{Two-phase Topology Recovery Algorithm}
\label{Alg1}
\end{algorithm}

\subsection{Augmented Lagrangian topology recovery}
\label{augmented}
In this subsection we develop a constrained independent component analysis  (cICA) method \cite{1388469}
to solve (\ref{optimization5}).
This approach is based on sequentially estimating the demixing matrix,
$\Wmat\define\tilde{\Lmat}^{-1}$, under constraints,
where
the inequality
constraints (Constraints 
$2)$ and  
$3)$ from \eqref{optimization5})
  are transformed into equality constraints
	in the  augmented Lagrangian {\cite{BERTSEKAS_1982,LU_Rajapakse_2006}}.
	Constraint 1) implies the symmetry of 
$\Wmat$, i.e. the equality constraint $\Wmat=\Wmat^T$.
Thus, in this case
the  objective function for the cICA,
which is based on
Equation (3) in \cite{1388469}
 is given by 
	\beqna
\label{augmented2}
Q_a(\Wmat,\muvec,\Lambdamat,\Dmat)
=-\psi\left(\Wmat^{-1}\right)\hspace{3cm}
\nonumber\\
+
\frac{1}{2\gamma}
\sum_{m=1}^{M-1}(\{-\gamma \sum_{l=1}^{M-1}\Wmat^{-1}_{m,l}+\mu_{m}\}^+)^2-\mu_m^2
\hspace{0.2cm}\nonumber\\
+\frac{1}{2\gamma}\sum_{m=1}^{M-1}\sum_{k=1}^{m-1} (\{{\gamma}\Wmat^{-1}_{m,k}
+\Lambdamat_{k,m}\}^+)^2-\Lambdamat_{k,m}^2
\nonumber\\
- \sum_{m=1}^{M-1}\sum_{k=1}^{m-1}\Dmat_{m,k}
(\Wmat_{m,k}-\Wmat_{k,m})
\hspace{1.6cm}
\nonumber\\+
\frac{\gamma}{2}\sum_{m=1}^{M-1}\sum_{k=1}^{m-1}(\Wmat_{m,k}-\Wmat_{k,m})^2,
\hspace{1.8cm}
\eeqna
where $\muvec$, $\Lambdamat\succeq \zerovec$, and $\Dmat$
are  the nonnegative vector,  positive semidefinite matrix, and symmetric matrix,
respectively, of  Lagrange
multipliers, and  $\gamma>0$   is the penalty
parameter.
The minimization of (\ref{augmented2})
w.r.t.  $\Wmat$ results in the following 
 natural gradient descent learning rule \cite{Amari_1998} for $\Wmat$:
\beqna
\label{WWW}
\Wmat^{(t+1)}=\Wmat^{(t)}
-\eta \nu\left(\Wmat^{(t)},\muvec^{(t+1)},\Lambdamat^{(t+1)},\Dmat^{(t+1)}\right),
\eeqna
where $t=0,1,\ldots$ is the iteration index,
\beqna
	\label{delta_w2}
	\nu\left(\Wmat,\muvec,\Lambdamat,\Dmat\right)
	\define
\Wmat^T\frac{\partial Q_a(\Wmat,\muvec,\Lambdamat,\Dmat)}{\partial \Wmat}\Wmat^T,
\eeqna
and $0<\eta\leq 1$
 is the  learning rate that determines the step size.
By substituting \eqref{cov_tilde} and $\Wmat=\tilde{\Lmat}^{-1}$
in (\ref{Gaussian22}) and then taking the derivative of the result w.r.t. $\Wmat$, we obtain
	\beqna
\label{likelihood1_der}
\frac{\partial \psi(\Wmat^{-1})}{\partial \Wmat}=\hspace{6.2cm}\nonumber\\
-\Wmat^{-T}\left(\hat{\bsigma}_{\tilde{\pvec}}-\hat{\sigma}^2\Umat^\dagger  (\Umat^\dagger)^T\right) \Wmat^{-1} \sigmatilde^{-1} \Wmat^{-T}
+\Wmat^{-T}.
\eeqna
By substituting \eqref{likelihood1_der}
in \eqref{delta_w2}, we obtain
\beqna
	\label{delta_w3}
	\nu\left(\Wmat,\muvec,\Lambdamat,\Dmat\right)
=
\left(\hat{\bsigma}_{\tilde{\pvec}}-\hat{\sigma}^2\Umat^\dagger  (\Umat^\dagger)^T\right)\Wmat^{-1} \sigmatilde^{-1} 
\nonumber\\
-\Wmat^{T}+
	 {\mathbf{1}}_{M-1}\muvec^T
	-\Lambdamat
-\Wmat^{T}(\Dmat^T-\Dmat)\Wmat^{T}.
	\eeqna
Finally, the Lagrange multipliers, $\muvec$, $\Lambdamat$, and $\Dmat$,  according to the gradient
ascent method are updated as follows:
\beqna
\label{uuu}
\muvec^{(t+1)}&=&\left\{\muvec^{(t)}-\gamma (\Wmat^{(t)})^{-1}{\mathbf{1}}_{M-1}\right\}^+,
\\
\label{lll}
\Lambdamat^{(t+1)}&=&\left\{\Lambdamat^{(t)} +\gamma {\text{off}}(\Wmat^{(t)})^{-1}  \right\}^+,
	\\
\label{rrr}
	\Dmat^{(t+1)}&=&\Dmat^{(t)}
	-\gamma \left(\Wmat^{(t)}-(\Wmat^{(t)})^T\right),
	\eeqna
$m,k=1,\ldots,M-1$.
$\Lambdamat^{(t+1)}$ is a symmetric matrix with nonnegative elements and 
zero diagonal. 
Then, it is updated according to \eqref{WWW}-\eqref{rrr} until convergence.


The augmented Lagrangian topology recovery  is summarized in Algorithm \ref{Alg3}.
The ML-BEST approach with augmented Lagrangian topology recovery is implemented by Algorithm \ref{Alg0}, where Step 5 is implemented by Algorithm \ref{Alg3}.
\begin{algorithm}[ht]
 \SetAlgoLined
{\bf{Input: }}
$\sigmatilde$, $\hat{\bsigma}_{{\pvec}}$, $\hat{\sigma}^2$.\\
 {\bf{Output:}} Estimator $\hat{\tilde{\Lmat}}^{\text{ML}}$.\\
{\bf{Algorithm Steps:}}
\begin{enumerate}
\item Evaluate  the reduced sample covariance matrix
  from (\ref{cov_est})
by $\hat{\bsigma}_{\tilde{\pvec}}=\Umat^\dagger \hat{\bsigma}_{{\pvec}} (\Umat^\dagger)^T$.
\item Initialize $\hat{\tilde{\Lmat}}^{(0)}$, for example, by the estimator from \eqref{H_estimation}:
\beqna
\hat{\tilde{\Lmat}}^{(0)}=\hat{\tilde{\Lmat}}^{{\text{PD}}}.
\nonumber
\eeqna
\item Set $t=0$, $\uvec^{(0)}=\zerovec$, $\Lambdamat^{(0)}=\zerovec$,  $\Wmat^{(0)}=\left(\hat{\tilde{\Lmat}}^{(0)}\right)^{-1}$,
and $\gamma,\eta>0$  to small positive scalar values.
\item {\bf{Repeat}}
\begin{enumerate}
\item Update
\[\Wmat^{(t+1)}=\Wmat^{(t)}-\eta \nu\left(\Wmat^{(t)},\muvecsmall^{(t+1)},\Lambdamat^{(t+1)},\Dmat^{(t+1)}\right),
\]
where $ \nu(\cdot)$ is given in \eqref{delta_w3}.
\item Update the Lagrange multipliers, 
$\uvec^{(t+1)}$, $\Lambdamat^{(t+1)}$, and $\Dmat^{(t+1)}$,
 according to (\ref{uuu}), (\ref{lll}), and (\ref{rrr}), respectively.
\item $t\rightarrow t+1$
\end{enumerate}
{\bf{Until}} criterion $||\Wmat^{(t+1)}-\Wmat^{(t)}||_F\leq  \epsilon$.
\item Evaluate the reduced topology matrix $\hat{\tilde{\Lmat}}=\left(\Wmat^{(t+1)}\right)^{-1}$.
\end{enumerate}
 \caption{Augmented Lagrangian Topology Recovery Algorithm}
\label{Alg3}
\end{algorithm}

\section{Remarks}
\label{remark_section}
In this section, we discus the identifiability conditions
and complexity in Subsection \ref{Ident_section} and
\ref{sec_complex}, respectively, and
describe a few extensions for the proposed model and methods in Subsection
\ref{extension_section}.
\subsection{Identifiability conditions}
\label{Ident_section}
In this subsection, we discuss the GBSS identifiability conditions,
under which the topology matrix and the state vectors can be recovered \cite{Tong_Huang_1991} for the model from Section \ref{problem_formulation_sec} with zero-mean measurements.
It is well known that Gaussian sources with i.i.d. time-structures cannot be separated  
\cite{Comon_1994,Pham_Garat_1997,Cardoso_1998}.
Nevertheless, the following theorem states that when the mixing matrix is a symmetric matrix,
 consistent separation can rely exclusively on the SOS of the source covariance,  
even for Gaussian sources.
\begin{theorem}
\label{Th1}
Given the model in \eqref{the_model_0} and the relation in \eqref{relation},
and assuming the following conditions:
\begin{itemize}
\item  $\tilde{\Lmat}$ is a symmetric positive definite matrix
\item The covariance of the states, $\sigmatilde$, is known and  is a positive definite matrix
\item The matrix $\hat{\bsigma}_{\tilde{\pvec}}-\hat{\sigma}^2 \Umat^\dagger  (\Umat^\dagger)^T
$, where $\hat{\bsigma}_{\tilde{\pvec}}$ and $\hat{\sigma}^2$ are defined in \eqref{cov_est} and \eqref{MLvariance}, respectively, is a positive semidefinite matrix.
\end{itemize}
Then, the Laplacian mixing
matrix, ${\Lmat}$, can be  uniquely identified,
without scaling and permutation ambiguities, from the sample covariance matrix of the observations, $\hat{\bsigma}_{\pvec}$, defined in \eqref{total_cov}.
\end{theorem}
\begin{IEEEproof}
First we will show that $\tilde{\Lmat}$ is identifiable.
Then, ${\Lmat}$ can be uniquely recovered  by using the relationship in \eqref{relation}.
Similar to the derivation of (\ref{cov_tilde}), it can be shown 
that for any state distribution and independent noise
with known noise covariance, $\sigma^2\Imat_M$,
 the covariance of the observations, $\tilde{\pvec}[n]$, $n=0,\ldots,N-1$, satisfies
\beqna
\label{cov_tilde_identi}
\bsigma_{\tilde{\pvec}}(\tilde{\Lmat},\sigma^2)
&=&\tilde{\Lmat}^{T}\sigmatilde \tilde{\Lmat} +{\sigma}^2 \Umat^\dagger  (\Umat^\dagger)^T
\nonumber\\
&=&\tilde{\Lmat}\sigmatilde \tilde{\Lmat} +{\sigma}^2 \Umat^\dagger  (\Umat^\dagger)^T
,
\eeqna
where the last equality is obtained by substituting the symmetry property, $\tilde{\Lmat}^{T}=\tilde{\Lmat}$.
It is known that for any positive definite matrix $\Amat\in{\mathbb{R}}^{(M-1)\times(M-1)}$ there exists a {\em{unique}} positive definite square root,  $\Bmat \in {\mathbb{R}}^{(M-1)\times(M-1)}$, such that $\Amat=\Bmat^2$ (see, e.g. p. 448 in \cite{Horn_Johnson_book}).
Thus, under the assumption that $\sigmatilde$ and $\bsigma_{\tilde{\pvec}}(\tilde{\Lmat},\sigma^2)-{\sigma}^2 \Umat^\dagger  (\Umat^\dagger)^T
$ are positive definite matrices,
the solution of \eqref{cov_tilde_identi} is {\em{unique}} and  is given by
\be
\label{est2}
\tilde{\Lmat}
=\sigmatilde^{-\frac{1}{2}}\left(\sigmatilde^{\frac{1}{2}}
\left(\bsigma_{\tilde{\pvec}}(\tilde{\Lmat},\sigma^2)-{\sigma}^2\Umat^\dagger  (\Umat^\dagger)^T\right)
\sigmatilde^{\frac{1}{2}}\right)^{\frac{1}{2}}\sigmatilde^{-\frac{1}{2}}.
\ee
Now, if we  use the estimators $\hat{\bsigma}_{\tilde{\pvec}}$ and $\hat{\sigma}^2$ in \eqref{est2} instead of the true unknown values of
${\bsigma}_{\tilde{\pvec}}(\tilde{\Lmat},\sigma^2)$, $\sigma^2$, then
the existence
of a positive definite solution is not guaranteed.
Under the Theorem's assumption
 that  $\hat{\bsigma}_{\tilde{\pvec}}-\hat {\sigma}^2 \Umat^\dagger  (\Umat^\dagger)^T$
is a positive definite matrix,
the uniqueness holds for the solution in \eqref{H_estimation}.
\end{IEEEproof}

A necessary condition 
for the existence of the inverse of $\hat{\bsigma}_{\tilde{\pvec}}$,  as required in Theorem \ref{Th1},
is that the sample covariance matrix has a full rank, i.e.
${\text{rank}}(\hat{\bsigma}_{\tilde{\pvec}}) = M-1$.
To ensure  numerical stability,
 we require stricter conditions than the condition $N\geq M-1$.
However, by using the sparsity assumption, this condition
can be relaxed even further.
When the mixing
matrix, $\tilde{\Lmat}$, is invertible, identifiability of the mixing matrix implies
the ability to separate the sources, for example, by the MMSE estimator, as shown in Subsection \ref{MMSE_subsection}.

\subsection{Complexity}
\label{sec_complex}
In this section we analyze the computational complexity of the  proposed ML-BEST methods,
based on the number of multiplications of the matrix operations. 
The multiplications and pseudo-inverse calculations of $\Umat$ from \eqref{AAA} are not taken into account, since they are not an inherent part of the algorithms.

\begin{enumerate}
\item {\em{Basic ML-BEST approach}}\\
Algorithm \ref{Alg0} shows the basic ML-BEST approach.
The computational complexity of the multiplication for calculating the sample covariance  matrix in Step 2
is  ${\cal{O}}(N M^2)$.
Then, finding the smallest eigenvalue of this matrix at Steps 3-4
calls for eigendecomposition
or matrix inversion of the $M\times M$ sample covariance matrix, each typically requiring computational complexity on the order of 
${\cal{O}}(M^3)$.
Thresholding the resultant Laplacian matrix estimator at Step 7 costs ${\cal{O}}(M^2)$.
Then,  the state estimation at Step 8
 costs ${\cal{O}}(3M^3+3M^3+N M^2)$, since it
requires the pseudo-inverse of an $M\times M$ matrix and 
3 multiplications of $M\times M$ matrices, in addition to $N$ times the multiplication of an $M$-length vector with a square matrix.
Thus, the total complexity of the ML-BEST algorithm (without the topology recovery step) is
${\cal{O}}((2N+1)M^2+7M^3)$.
\item {\em{Two-phase topology recovery}}\\
Algorithm \ref{Alg1} shows the two-phase topology recovery algorithm.
 The complexity of calculating $\hat{\tilde{\Lmat}}^{{\text{PD}}}$
at Step 2 consists of calculating the singular value decomposition (SVD) of an $M\times M$ matrix in order to obtain its square root,
and 4 multiplications of $M\times M$ matrices and, thus, it costs 
${\cal{O}}(5M^3)$.
The nonnegative quadratic program in Step 3 has polynomial time solutions,
where its exact computational complexity 
 depends on the solver, method, and exact problem parameters.
Here,
we approximate this polynomial complexity by ${\cal{O}}(P^2 K)$, where $P$ is the  number of real decision variables and
 $K$ is the number of constraints.
In our case, we have $P=\frac{M(M-1)}{2}$ scalar real
decision variables and 
\beqna K=\underbrace{M}_{{\text{pos. diag}}}+
\underbrace{\frac{(M-1)(M-2)}{2}}_{{\text{neg. off-diag}}}+\underbrace{M-1}_{{\text{diag dom.}}}=\frac{M(M+1)}{2}\nonumber
\eeqna
linear constraints  on these variables that  stem from Constraints  $1)-3)$ in (\ref{optimization_after_ML}).
Thus, the computational complexity of Step 3 is around
${\cal{O}}(M^3(M^3-M^2-M+1))$,
and the total complexity of the two-phase topology recovery algorithm  is
${\cal{O}}(M^3(M^3-M^2-M+1))$.
\item{\em{Augmented Lagrangian topology recovery}}\\
Algorithm \ref{Alg3} shows the augmented Lagrangian topology recovery algorithm.
The complexity of the initialization step depends on the selected initial estimator.
If, for example, we initialize with $\hat{\tilde{\Lmat}}^{{\text{PD}}}$, then  it costs 
${\cal{O}}(5M^3)$, as explained in the previous algorithm.
For each iteration the computational complexity of Step 4.a is based on $M\times M$ matrix multiplications and inversion,
which costs ${\cal{O}}(5M^3)$.
The complexity of Step 4.b of calculating the Lagrange multipliers by the thresholding operator (versus zero) is of order 
${\cal{O}}(5M^3)$.
Typically, it takes $100-1000$ iterations to converge.
\end{enumerate}

Based on the above exposition, the computational complexities of Algorithms \ref{Alg1} and \ref{Alg3} for topology recovery are of the order ${\cal{O}}(M^3(M^3-M^2-M+1))$ and ${\cal{O}}(M^3)$, respectively.
Thus, if we were to let $M$ grow
while keeping $N$ fixed, the  augmented Lagrangian topology recovery method would 
exhibit significant computational savings when
compared to the two-phase topology recovery.

\subsection{Possible extensions}
\label{extension_section}
\subsubsection{Extension for general states distribution}
In the case where the states are non-Gaussian, 
we can develop the constrained ML-BEST similarly to the derivations in Section \ref{ML_BEST_section}.
That is, 
we assume the  model from (\ref{General_graph3}) and compute the reduced source pdf, 
 $f_{\tilde{\thetavecsmall}}(\cdot)$, by using a
transformation of pdf rules  (see, e.g. pp. 134-144 \cite{Papoulis}).
Under these assumptions, 
the normalized log likelihood  of $\tilde{\pvec}[n]$, $n=0,\ldots,N-1$,
is given by \cite{Cardoso_1998}
\beqna
\label{likelihood1}
\psi(\tilde{\Lmat})
=\frac{1}{N}\sum_{n=0}^{N-1} 
\log f_{\tilde{\thetavecsmall}}\left( \tilde{\Lmat}^{-1}\tilde{\pvec}[n]\right)
-\log |\tilde{\Lmat}|.
\eeqna
Then, the ML is obtained by minimizing (\ref{likelihood1}) under 
 the reduced-Laplacian matrix
 Properties P.1-P.5, similarly to in the problem formulated in 
(\ref{optimization5}).
If direct KKT solution of this constrained minimization is intractable,
we can develop associated  low-complexity methods,  similarly to in Subsections \ref{two_phase_CML} and \ref{augmented}.

Alternatively the proposed Gaussian ML-BEST methods  can also be applied for non-Gaussian distributions  with the same covariance, since the structure of the covariance matrices in \eqref{cov_p} and \eqref{cov_tilde} holds for any distribution. Although this ML-BEST approach may not be optimal for non-Gaussian distributions, it has the advantage of only requiring the SOS. In addition, SOS methods are expected to be more robust in adverse SNRs \cite{Belouchrani_Cardoso_Moulines_1997}.
\subsubsection{Shunt in admittance matrix}
In many cases, the bus admittance matrix contains a shunt, representing the bus admittance-to-ground connection. 
Shunt elements are not considered here; nevertheless, the proposed model and methods can be easily
extended to the case of some shunt elements by  adding
 the shunt  elements to the diagonal terms of the matrix $\Lmat$. In this case, the symmetry of the matrix $\Lmat$ is preserved, but $\Lmat$ becomes a nonsingular matrix and the assumption of a reference bus is redundant.

\section{CRB}
\label{CRB_section}
The CRB is a commonly-used  lower bound on the mean-squared error (MSE) matrix of any unbiased estimator of a deterministic parameters vector. 
In this section, we derive a  closed-form expression for the CRB of the
 mixing Laplacian matrix and the noise variance, by modeling the sources as  nuisance random parameters.

By using the symmetry of the matrix $\tilde{\Lmat}$,
we define
the vector of unknown parameters for the CRB as
\[\alphavec\define[{\text{vech}}(\tilde{\Lmat})^T,\sigma^2]^T\in{\mathbb{R}}^{\frac{M(M-1)}{2}+1},\]
which
 consists
 of the lower triangular elements of $\Lmat$, including the diagonal, and the noise variance, $\sigma^2$.
Then, under some mild regularity condition, the CRB on the MSE of any unbiased estimator of
$\alphavec$ is given by
\beqna
\label{bound1}
{\rm{E}}\left[\left(\hat{\alphavec}-\alphavec\right)\left(\hat{\alphavec}-\alphavec\right)^T\right]
\succeq \Bmat_{\text{CRB}}(\alphavec)=
\Jmat^{-1}(\alphavec),
\eeqna
where $\Jmat(\alphavec)$ is the associated Fisher information matrix (FIM).

To compute the CRB,  we stack the measurements $\tilde{\pvec}[n]$
from \eqref{General_graph4} into a  single $(M-1)N$ length  vector, such that
\be
\label{General_graph4_vec}
\tilde{\pvec}=(\Imat_N\otimes\tilde{\Lmat}) \tilde{\thetavec}+\tilde{\wvec},
\ee
where
$\tilde{\pvec}\define[\tilde{\pvec}^T[0],\ldots,\tilde{\pvec}^T[N-1]]^T$,
$\tilde{\thetavec}\define[\tilde{\thetavec}^T[0],\ldots,\tilde{\thetavec}^T[N-1]]^T$,
and
$\tilde{\wvec}\define[\tilde{\wvec}^T[0],\ldots,\tilde{\wvec}^T[N-1]]^T$.
According to the model assumptions, $\tilde{\thetavec}$
and, therefore, also 
$\tilde{\pvec}$ are zero-mean random vectors. Under the assumption of time-independent states, the $(M-1)N\times(M-1)N$ covariance
matrix of $\tilde{\thetavec}$ is a block-diagonal matrix with the structure
\beqna
\label{Ctheta}
\Cmat_{\tilde{\thetavecsmall}}\define {\rm{E}}[\tilde{\thetavec}\tilde{\thetavec}^T ]=
(\Imat_N \otimes \sigmatilde).
\eeqna
Thus, the $(M-1)N\times(M-1)N$ covariance matrix of $\tilde{\pvec}$,
$\Cmat_{\tilde{\pvec}}\define {\rm{E}}[\tilde{\pvec}\tilde{\pvec}^T ]$,  is  given
by
\beqna
\label{cmat}
\Cmat_{\tilde{\pvec}}
&=&(\Imat_N\otimes\tilde{\Lmat}^T)\Cmat_{\tilde{\thetavecsmall}}(\Imat_N\otimes\tilde{\Lmat})
+(\Imat_N\otimes\sigma^2 \Umat^\dagger  (\Umat^\dagger)^T)
\nonumber\\
&=&\left(\Imat_N\otimes
\left(\tilde{\Lmat} \sigmatilde \tilde{\Lmat}
+\sigma^2 \Umat^\dagger  (\Umat^\dagger)^T\right)\right),
\eeqna
where the last equality is obtained by substituting \eqref{Ctheta} with $\tilde{\Lmat}^T=\tilde{\Lmat}$, and using 
Kronecker product associativity and the rule, $(\Amat_1\otimes \Amat_2)(\Amat_3 \otimes\Amat_4)=(\Amat_1\Amat_3 \otimes \Amat_2\Amat_4)$ for any set of matrices $\Amat_i$, $i=1,2,3,4$ with appropriate dimensions.

Due to the zero-mean Gaussian distribution of $\tilde{\pvec}$, the  $(m,r)$ entry of the associated $\left(\frac{M(M-1)}{2}+1\right)\times\left(\frac{M(M-1)}{2}+1\right)$ FIM is given by (see, e.g. p. 48 in \cite{Kayestimation})
\beqna
\label{Jklpq}
\Jmat_{m,r}(\alphavec)=
\frac{1}{2}{\text{Tr}}\left\{\Cmat_{\tilde{\pvec}}^{-1}\frac{\partial \Cmat_{\tilde{\pvec}}}{\partial \alpha_{m} }\Cmat_{\tilde{\pvec}}^{-1} \frac{\partial \Cmat_{\tilde{\pvec}}}{\partial 
\alpha_{r}} \right\},
\eeqna
for any $m,r=1,\ldots,\frac{M(M-1)}{2}+1$.
The derivatives of \eqref{cmat} w.r.t. the elements of $\tilde{\Lmat}$ and $\sigma^2$  are given by
\beqna
\label{diffL}
\frac{\partial \Cmat_{\tilde{\pvec}}}{\partial\alpha_r }=
\frac{\partial \Cmat_{\tilde{\pvec}}}{\partial \tilde{\Lmat}_{k,l} }
=
\left(\Imat_N
\otimes \frac{\partial \tilde{\Lmat}^T \sigmatilde \tilde{\Lmat}}{\partial \tilde{\Lmat}_{k,l} } \right)
=
\left(1-\frac{1}{2}\delta_{k,l}\right)
\nonumber\\
\times\left(\Imat_N
\otimes \left( (\Emat_{k,l}+\Emat_{l,k}) \sigmatilde\tilde{\Lmat}+\tilde{\Lmat}\sigmatilde(\Emat_{k,l}+\Emat_{l,k})
\right)\right),
\eeqna
where $\Emat_{k,l}=\evec_l\evec_l^T$, and for $r=1,\ldots, \frac{M(M-1)}{2}$, and where $r$ is such that $\alpha_r=\tilde{\Lmat}_{k,l}$, and
\beqna
\label{diffSigma}
\frac{\partial \Cmat_{\tilde{\pvec}}}{\partial\alpha_r }=
\frac{\partial \Cmat_{\tilde{\pvec}}}{\partial \sigma^2}
=
 \left(\Imat_N
\otimes \Umat^\dagger  (\Umat^\dagger)^T\right),
\eeqna
for $r={\frac{M(M-1)}{2}}+1$.

By substituting \eqref{cmat} and \eqref{diffL} in \eqref{Jklpq},
using Kronecker product rules, the symmetry of the matrices, and the trace  operator rule,
 ${\text{Tr}}\{(\Amat\otimes\Bmat)\}={\text{Tr}}\{\Amat\}{\text{Tr}}\{\Bmat\}$,
one obtains
\beqna
\label{Jklpq21}
\Jmat_{m,r}(\alphavec)
=
\frac{N}{2}\left(1-\frac{1}{2}\delta_{k,l}\right)\left(1-\frac{1}{2}\delta_{q,p}\right)
\hspace{1cm}\nonumber\\
\times{\text{Tr}}\left\{
\left(\tilde{\Lmat} \sigmatilde \tilde{\Lmat}+ \sigma^2\Umat^\dagger  (\Umat^\dagger)^T\right)^{-1} 
\right.\hspace{2cm}
\nonumber\\
 \times \left( (\Emat_{k,l}+\Emat_{l,k}) \sigmatilde\tilde{\Lmat}+\tilde{\Lmat}\sigmatilde(\Emat_{k,l}+\Emat_{l,k})
\right)\hspace{0.5cm}
\nonumber\\
\times
\left(\tilde{\Lmat} \sigmatilde \tilde{\Lmat}+ \sigma^2 \Umat^\dagger  (\Umat^\dagger)^T\right)^{-1}\hspace{2.7cm}
\nonumber\\
\left. \times
  \left( (\Emat_{p,q}+\Emat_{q,p}) \sigmatilde\tilde{\Lmat}+\tilde{\Lmat}\sigmatilde(\Emat_{p,q}+\Emat_{q,p}) \right)\right\},
\eeqna
 $\forall m,r=1,\ldots,\frac{M(M-1)}{2}$,
where $m$ and $r$ are such that $\alpha_m=\tilde{\Lmat}_{k,l}$ and  $\alpha_r=\tilde{\Lmat}_{p,q}$.
Thus, 
$\Jmat_{m,r}(\alphavec)$ from \eqref{Jklpq21} includes the mutual FIM between the
elements of the lower triangular of the mixing matrix, $\tilde{\Lmat}_{k,l}$ and  $\tilde{\Lmat}_{p,q}$,
such that
 $ k,l,p,q=1,\ldots,M-1$, $l\leq k$, $p\leq q$.
By using the trace and vectorization operators rule, it can be verified that
\beqna
\label{A1234}
{\text{Tr}}\{(\Amat_1\Amat_2)^T\Amat_2\Amat_3\}
=
({\text{vec}}(\Amat_1))^T(\Amat_2\otimes\Amat_2){\text{vec}}(\Amat_3),
\eeqna
for any set of matrices $\Amat_i$, $i=1,2,3$  of compatible dimensions.
By applying \eqref{A1234} on \eqref{Jklpq21}
with the matrices
\[\Amat_1= (\Emat_{k,l}+\Emat_{l,k}) \sigmatilde\tilde{\Lmat}+\tilde{\Lmat}\sigmatilde(\Emat_{k,l}+\Emat_{l,k}),\]
\[\Amat_2=\left(\tilde{\Lmat} \sigmatilde \tilde{\Lmat}+ \sigma^2\Umat^\dagger  (\Umat^\dagger)^T\right)^{-1},\]
and
\[\Amat_3= (\Emat_{p,q}+\Emat_{q,p}) \sigmatilde\tilde{\Lmat}+\tilde{\Lmat}\sigmatilde(\Emat_{p,q}+\Emat_{q,p}),\]
and using the symmetry of these matrices, 
the $(m,r)$ entry of the FIM from \eqref{Jklpq21} can be rewritten as
\beqna
\label{Jklpq2}
\Jmat_{m,r}(\alphavec)
=\frac{N}{2}\psivec^T(l,k)\Qmat\psivec(p,q),
\eeqna
$\forall m,r=1,\ldots,\frac{M(M-1)}{2}$, where $m$ and $r$ are such that $\alpha_m=\tilde{\Lmat}_{k,l}$ and  $\alpha_r=\tilde{\Lmat}_{p,q}$,
and where
\beqna
\label{Gamma_def}
\Qmat\define\hspace{7.5cm}
\nonumber\\ \left(\tilde{\Lmat} \sigmatilde \tilde{\Lmat}+ \sigma^2\Umat^\dagger  (\Umat^\dagger)^T\right)^{-1}\otimes
\left(\tilde{\Lmat} \sigmatilde \tilde{\Lmat}+ \sigma^2\Umat^\dagger  (\Umat^\dagger)^T\right)^{-1}
\eeqna
and
\beqna
\label{psi_def}
\psivec (l,k)\define \left(1-\frac{1}{2}\delta_{k,l}\right)
\hspace{4.5cm}
\nonumber\\
\times{\text{vec}}\left( (\Emat_{k,l}+\Emat_{l,k}) \sigmatilde\tilde{\Lmat}+\tilde{\Lmat}\sigmatilde(\Emat_{k,l}+\Emat_{l,k})
\right).
\eeqna

Similarly, by substituting \eqref{cmat}, \eqref{diffL}, and \eqref{diffSigma} in \eqref{Jklpq},
and using the symmetry of the matrices,
we obtain that the  $(m,r)$ entry of the  FIM is
\beqna
\label{Jkls2}
\Jmat_{m,s}(\alphavec)=\frac{N}{2}
\psivec^T(l,k)\Qmat
{\text{vec}}(\Umat^\dagger(\Umat^\dagger)^T),\hspace{1.3cm}
\\
\label{Jskl2}
\Jmat_{s,m}(\alphavec)=\frac{N}{2}
({\text{vec}}( \Umat^\dagger(\Umat^\dagger)^T))^T \Qmat\psivec(l,k)
,\hspace{1cm}
\\
\label{Jss2}
\Jmat_{s,s}(\alphavec)=\frac{N}{2}({\text{vec}}(\Umat^\dagger(\Umat^\dagger)^T))^T\Qmat{\text{vec}}(\Umat^\dagger(\Umat^\dagger)^T),
\eeqna
for $s=\frac{M(M-1)}{2}+1$, $m=1,\ldots,\frac{M(M-1)}{2}$, and
$m$ is such that $\alpha_m=\tilde{\Lmat}_{k,l}$.

Equations \eqref{Jklpq2} and \eqref{Jkls2}-\eqref{Jss2} imply that  the FIM
can be formulated in a matrix form as follows:
\beqna
\label{J_final}
\Jmat(\alphavec)=\frac{N}{2}\Psimat^T\Qmat\Psimat,
\eeqna
where the matrix $\Psimat$ is an $ (M-1)^2\times\left(\frac{M(M-1)}{2}+1\right)$ matrix in which the
first $\frac{M(M-1)}{2}$  columns 
are the vectors $\psivec(l,k)$ ordered with the same order as ${\text{vech}}(\tilde{\Lmat})$
and the last column is ${\text{vec}}(\Umat^\dagger(\Umat^\dagger)^T)$.

By substituting \eqref{J_final} in \eqref{bound1} we obtain the CRB:
\beqna
\label{bound2}
{\rm{E}}\left[\left(\hat{\alphavec}-\alphavec\right)\left(\hat{\alphavec}-\alphavec\right)^T\right]
\succeq \Bmat_{\text{CRB}}(\alphavec)=\frac{2}{N}\left(\Psimat^T\Qmat\Psimat\right)^\dagger.
\eeqna
The bound from \eqref{bound2} implies, in particular,
 the lower bound on the MSE matrix of the lower triangular 
of the reduced-Laplacian matrix:
\beqna
{\rm{E}}\left[\left({\text{vech}}(\hat{\tilde{\Lmat}})-{\text{vech}}({\tilde{\Lmat}})\right)\left({\text{vech}}(\hat{\tilde{\Lmat}})-{\text{vech}}({\tilde{\Lmat}})\right)^T\right]
\nonumber\\
\succeq
\left[\Bmat_{\text{CRB}}(\alphavec)\right]_{1:\frac{M(M-1)}{2},1:\frac{M(M-1)}{2}}.
\eeqna
Similarly, the CRB on the noise variance is given by
\beqna
{\rm{E}}\left[(\hat{\sigma}^2-{\sigma}^2)^2\right]
\geq
\left[\Bmat_{\text{CRB}}(\alphavec)\right]_{\frac{M(M-1)}{2}+1,\frac{M(M-1)}{2}+1}.
\eeqna

To get more insight into \eqref{J_final}, we investigate the trivial case
of $\sigma^2=0$ and $\tilde{\Lmat}=c\Imat$, for $c>0$.
By substituting these values in \eqref{Jklpq21} and using the trace properties and the symmetry of the involved matrices, it can be verified that the  $(m,r)$ entry of the  FIM  in this case is
\beqna
\label{J_insight}
\Jmat_{m,r}(\alphavec)
=
\frac{N}{ 2c^2}{\text{Tr}}\left(1-\frac{1}{2}\delta_{k,l}\right)\left(1-\frac{1}{2}\delta_{q,p}\right)\hspace{0.6cm}
\nonumber\\
\times\left\{
(\Emat_{k,l}+\Emat_{l,k})(\Emat_{p,q}+\Emat_{q,p})
\right.\hspace{1cm}
\nonumber\\
+\left. \sigmatilde^{-1} (\Emat_{k,l}+\Emat_{l,k}) \sigmatilde(\Emat_{p,q}+\Emat_{q,p})
\right\},
\eeqna
where $m$ and $r$ are such that $\alpha_m=\tilde{\Lmat}_{k,l}$ and  $\alpha_r=\tilde{\Lmat}_{p,q}$.
Thus,  \eqref{J_insight} implies that
the elements of the FIM are nonzero in this case only
if $k=p$ and/or $k=q$ and/or $l=p$ and/or $l=q$.
That is, only if $\tilde{\Lmat}_{k,l}$ and  $\tilde{\Lmat}_{p,q}$
share a joint row or column in the Laplacian matrix. In terms of graphs, that means that 
only the connected
nodes influence the information for estimation.

In general BSS problems,  the CRB cannot be calculated 
and the induced CRB has been proposed as an alternative  \cite{Tichavsky_Oja_2006,Doron_Yeredor_Tichavsky_2007,Yeredor_2010,Anderson_Adali_2014}.
Here, due to the symmetry of the mixing matrix, we can obtain the associated CRB
from \eqref{bound2}.
Alternatively, this bound could be derived via the constrained CRB (CCRB) approach (see, e.g. \cite{Hero_constraint,Stoica_Ng,Nitzan_constraints}).
It should be  emphasized that in the evaluation of the CRB, which is a local bound, the inequality constraints do not contribute any side information \cite{Hero_constraint,Stoica_Ng,Nitzan_constraints} 
and  the sparsity constraint also does not affect the CRB
if the exact sparsity level is unknown \cite{sparse_con}.
Since
the only equality parametric constraint on the estimated Laplacian matrix in optimization in \eqref{optimization5}  is the symmetric constraint, it is the only constraint that is taken into consideration in the proposed graph CRB.

\section{Simulations}
\label{simulations_section}
In this section, we present simulation examples conducted in order to evaluate the performance
of the proposed ML-BEST methods from Algorithm \ref{Alg0}, combined with two-phase topology recovery and  with augmented Lagrangian topology recovery
from Algorithms \ref{Alg1} and \ref{Alg3},  respectively.
The optimization problems are solved using the CVX toolbox \cite{CVX},
the sparsity threshold is set according to  \eqref{tau}
with $\alpha=4/M$, and the step sizes,
$\eta$ and $\gamma$ in Algorithm \ref{Alg3} are tuned experimentally.
The simulations include two scenarios: IEEE 14-bus system \cite{data}
and  a random topology graph, with
 250 Monte-Carlo simulations for each scenario.
The MSE performance of the state estimators is compared with that of the
oracle MMSE estimator from \eqref{est_singular}.
In addition to the MSE of the vectorized topology estimators, ${\text{vech}} (\hat{\Lmat})$, the topology estimation performance is measured also by 
the F-score metric \cite{Egilmez_Pavez_Ortega_2016}:
\[
FS(\hat{\Lmat},\Lmat)\define\frac{2 tp}{2tp+fn+fp},
\]
where $tp$, $fp$, and $fn$ are the   true-positive, false-positive, and false-negative detection of graph
edges in $\hat{\Lmat}$  with respect to the ground truth edges in $\Lmat$.
The F-score takes values between $0$ and $1$, where the value $1$
means perfect classification. The F-score is a measure for the
error probability  in the connectivity matrix.
In addition, 
we use
the CRB from \eqref{bound2} as a benchmark in the experiments. 

\subsection{IEEE 14-bus power system}
In this subsection, we implement the proposed methods 
  for the IEEE 14-bus system, representing a portion of a power system in the Midwestern U.S.  The system parameters, such as branch
susceptances,   are taken from \cite{data} and  $M=14$.
The power flow measurements are generated using (\ref{the_model_0}).
The state covariance matrix is set to $\bsigma_\thetavecsmall= c^2\Imat_M$.
The SNR is defined as
$
{\text{SNR}}=\frac{1}{\sigma^2}{\text{Tr}}\left\{\tilde{\Lmat}\bsigma_{\tilde{\thetavecsmall}}\tilde{\Lmat}\right\}$.

\begin{figure}
        \begin{center}
          \begin{tabular}[t]{ccc}
					\hspace{-0.25cm}
            \subfigure{\psfig{figure=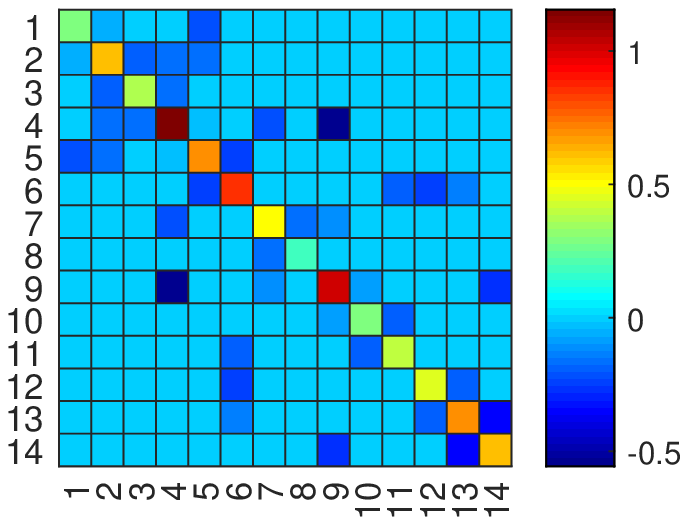,width=3cm}}	
							&\hspace{-0.25cm}
            \subfigure{\psfig{figure=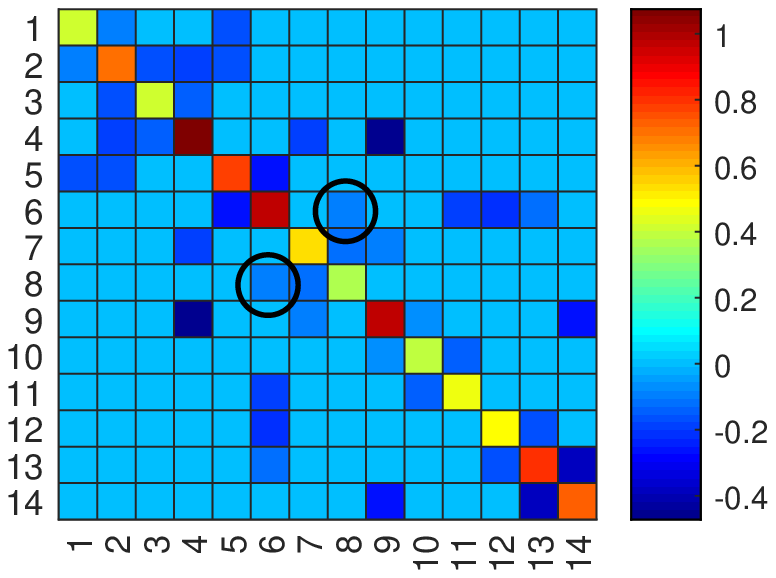,width=3cm}}
						&\hspace{-0.25cm}
            \subfigure{\psfig{figure=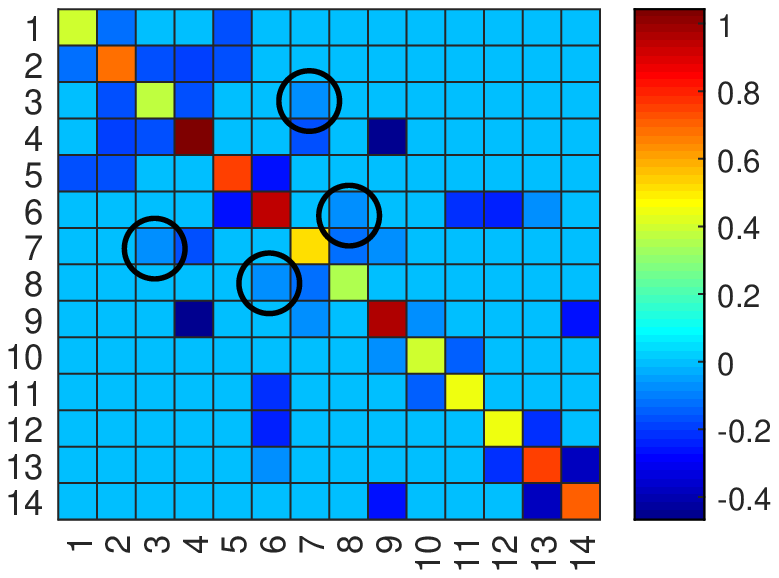,width=3cm}}
						    \end{tabular}
        \end{center}
				\caption{Illustration of the ML-BEST topology recovery methods to estimate the Laplacian matrix of the IEEE-14 bus system with $N=200$ samples and SNR$=15$dB:
				(a) the original Laplacian matrix;
				(b) and (c) the estimated Laplacian  by
  two-phase ML-BEST and Augmented Lagrangian ML-BEST methods, respectively.
	The black circles indicate false connections.}
\label{Tetris_fig1}
      \end{figure}

We first show in Fig. \ref{Tetris_fig1} visual comparisons between the Laplacian matrix of the IEEE 14-bus system and the associated  estimators of the Laplacian matrix, $\hat{\Lmat}$, obtained by the  two-phase ML-BEST and augmented Laplacian ML-BEST for $N=200$ and SNR$=15$dB.
The black circles in this figure indicate wrong connection estimation.
This comparison  shows that the positions of the
 entries in the estimated Laplacian matrices generally
correspond to the positions of the edges in the original
graph and, thus, the network could be constructed by the proposed procedures. 
Comparison between the two-phase ML-BEST  in (b) and the augmented Lagrangian ML-BEST
in (c) shows that the two-phase ML-BEST is better in terms of
support recovery. 
For example, while
 both methods identifies a false connection between bus 6 and bus 8,
only the augmented Lagrangian ML-BEST  identifies a false connection between bus 3 and bus 7.

\begin{figure}
        \begin{center}
          \begin{tabular}[t]{c}
            \subfigure[MSE of the topology estimation and the associated CRB]{\psfig{figure=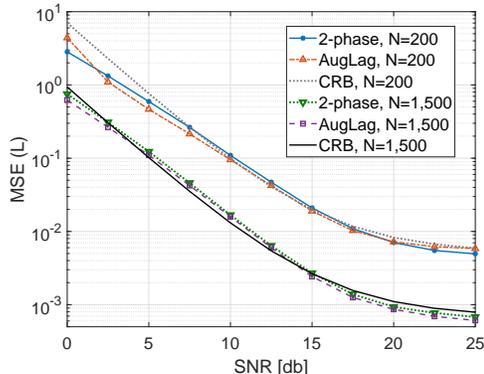,width=7cm}}		
          \\
            \subfigure[FS]{\psfig{figure=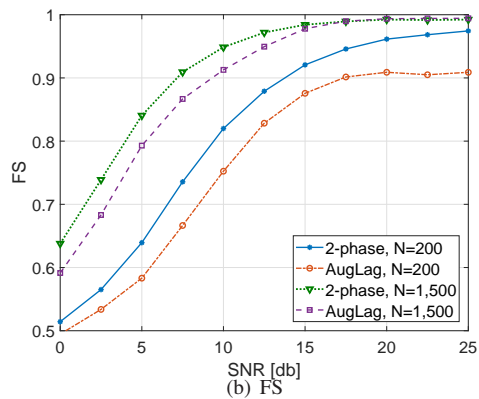,width=7cm}}
						\\
            \subfigure[MSE of the state estimators]{\psfig{figure=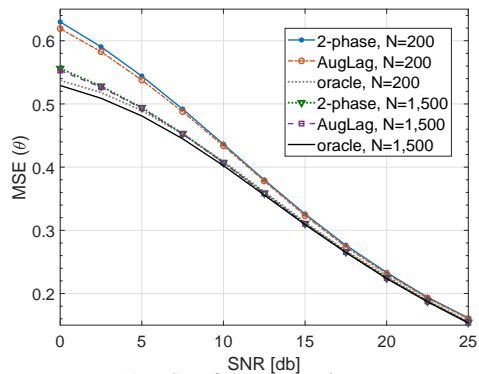,width=7cm}} 	
          \end{tabular}
        \end{center}
				\caption{The performance of the ML-BEST methods, with two-phase and augmented Lagrangian topology recovery,  for IEEE-14 bus system versus SNR with $N=200$, $1,500$.}
\label{Fig2}
      \end{figure}

The performance of the different methods is presented
in Fig. \ref{Fig2}   versus SNR
for $N=200$ and $N=1,500$.
It can be seen that the performance improves in any sense as $N$ increases, as expected.
In Fig \ref{Fig2}.a the MSE of the proposed ML-BEST methods for topology estimation and the associated CRB
are presented, and in Fig. \ref{Fig2}.b the F-score metric of the two ML-BEST methods is presented. 
It can be seen that while the two-phase topology recovery performs better in terms of F-score, the two ML-BEST methods have similar performance in terms of
MSE.
That is, the two-phase topology recovery is better in terms of estimating the connectivity matrix, i.e. it distinguishes between existing  and absent links,  
while the performance of both topology recovery methods  are close to the CRB for high SNR. However,
 since the CRB does not take into account the information on inequality constrains \cite{Hero_constraint,Stoica_Ng,Nitzan_constraints,sparse_con}, and especially the sparsity constraint,
it could be slightly higher than the true performance. 
The MSE of  the state estimators presented in Fig. \ref{Fig2}.c is similar for the two methods in this case. It can be seen  that 
for high SNRs, the state estimation performance of the ML-BEST methods with estimated topology converges to that of the oracle method, which uses the true topology.
Therefore, we can conclude that for high SNRs the topology estimation convergences to the true topology.

\subsection{Random topology}
In this subsection we simulate graphs from the Watts-Strogatz 'small world' graph model \cite{Watts_Strogatz} with  varying numbers of buses, $M$, and an average nodal degree  of 4,
which is shown to be appropriate for the simulation of synthetic power grid data \cite{Wang_Scaglione_Thomas_2010}.
It should be noted that the average nodal degree of a power network is almost invariant to the size of the network and, thus,
the sparsity level is usually a constant around
 $\frac{4M}{M^2}$.
The state covariance matrix set to $\bsigma_\thetavecsmall= c^2\Imat_M$,
with $c=\sqrt{10}$.
In order to achieve uniform SNR simulations, we set the norm of the Laplacian matrix to
$||\Lmat||_F=5$.

\begin{figure}
        \begin{center}
          \begin{tabular}[t]{c}
            \subfigure{\psfig{figure=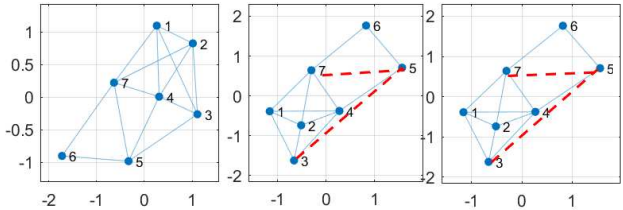,width=8.5cm}}	
									    \end{tabular}
        \end{center}
				\caption{Illustration of the ML-BEST topology recovery methods to estimate the Laplacian matrix of random topology with $N=200$ and $\sigma^2=1$:
				the original graph (left) and the estimated graph topology by
  two-phase ML-BEST (middle) and Augmented Lagrangian ML-BEST (right).
	The red lines indicate missing recovered edges.}
\label{graph_fig}
      \end{figure}

 Fig. \ref{graph_fig} presents a random graph and its recovery
by the  ML-BEST methods for $N=200$ and $\sigma^2=1$.
The red lines in this figure indicate missing connected edges
between buses 3 and 5, and buses 5 and 7.
This comparison  shows that the  estimated graphs are generally
correspond to the original
graph and, thus, the network could be reconstructed by the proposed procedures. 
The missing  recovered edges can be reconstructed by acquiring more data 
or by setting the sparse threshold more carefully.

\begin{figure}
        \begin{center}
          \begin{tabular}[t]{c}
            \subfigure[MSE of the topology estimation and the associated CRB]{\psfig{figure=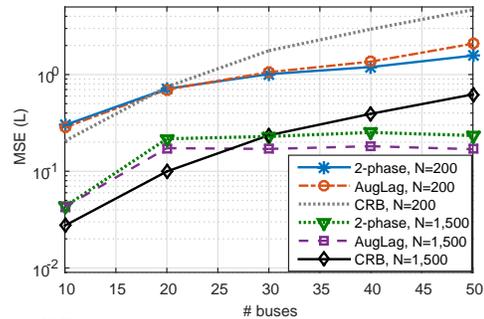,width=7cm}}			
          \\
            \subfigure[FS]{\psfig{figure=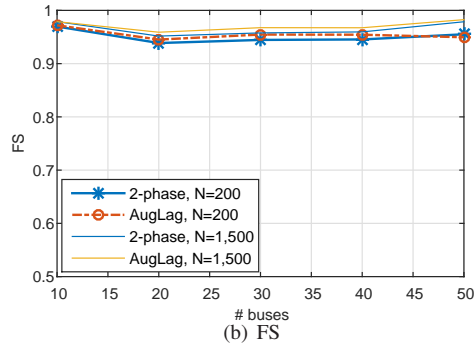,width=7cm}}
						\\
            \subfigure[MSE of the state estimators]{\psfig{figure=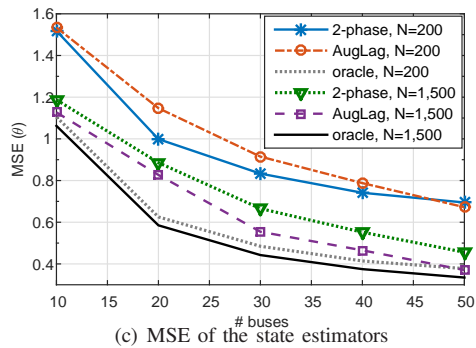,width=7cm}} 	
          \end{tabular}
        \end{center}
				\caption{The performance of the ML-BEST methods, with two-phase and augmented Lagrangian topology recovery,   for random topology versus the number of buses with $N=200,1,500$ and for $\sigma^2=1$.}
\label{Fig2_random}
      \end{figure}

The performance of the different methods for this random topology is presented
in Fig. \ref{Fig2_random}   
versus the number of buses in the system for $\sigma^2=1$, $N=200,1,500$,
and $c=\sqrt{10}$.
In Fig \ref{Fig2_random}.a the MSE of the  ML-BEST methods for topology estimation and the associated CRB
are presented. It can be seen that the topology MSE   degraded  as $M$ increases
since there are more parameters to estimate.
The CRB  does not take into account the sparsity and, thus, is higher than the true performance. However, it is still  a good predictor for the performance, and, thus, can be used for future system design.
In Fig. \ref{Fig2_random}.b the F-score metric of the two ML-BEST methods is presented. 
It can be seen that it is almost independent of the number of buses,
$M$, and that the two methods achieve similar results. 
The MSE of  the state estimators presented in Fig. \ref{Fig2_random}.c is lower for the two-phase ML-BEST method
with $N=200$, but for $N=1,500$ the Augmented Lagrangian ML-BEST has lower MSE.
Thus, different method should be used, depends of the number of samples.
The performance of the two methods become closer to those of the oracle performance as $N$ increases.
Since the mixing matrix has the same norm for any number of buses, $M$, the SNR is a constant. 
The structure of the Laplacian matrix leads to a lower MSE
of the state estimation as $M$ increases.

\begin{figure}[htb]
\centerline
{\psfig{figure=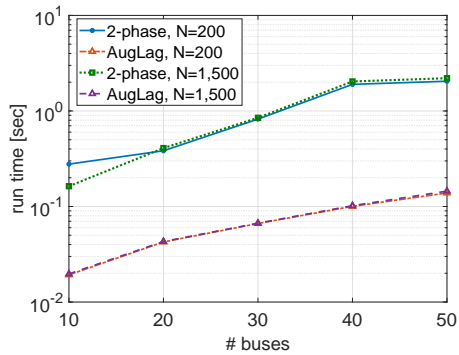,width=6.5cm}}
\caption{Runtime of the ML-BEST methods versus number of buses, $M$, in random topology with  $N=200,1,500$ samples.
}
\label{Fig10}
\vspace{-0.25cm}
\end{figure}

In order to demonstrate the empirical complexity of  the proposed methods  for different problem dimensions, the average computation  time, ``runtime", was evaluated by running the algorithm using Matlab on 
an Intel Core(TM) i7-7600U CPU computer, 2.80 GHz.
Figure \ref{Fig10} shows the runtime of the ML-BEST methods as a function of the
number of buses, $M$, 
for a random topology  and $N=200,1,500$ samples.
It can be seen that the runtime increases polynomially with the  number of buses, $M$, and it is higher for the two-phase topology recovery than for the augmented Lagrangian topology recovery with $100$ iterations,
 as expected from the theoretical discussion on computational complexity in Subsection \ref{sec_complex}.
The reason for this  is that the two-phase topology recovery stage from Algorithm \ref{Alg1}  requires solving an SDP
problem in \eqref{optimization_after_ML}  and, therefore, has a much higher computational
complexity as compared to the augmented ML-BEST estimator.
The number of measurements, $N$, has a less significant effect since it is only associated with the  cost of
computing the sample covariance matrix  and the state estimation at the beginning and the end of the basic ML-BEST approach.

%
%


%
\section{Conclusion}
\label{diss}
In this paper, we introduce the novel ML-BEST method for blind estimation of states and topology in power systems, by formulating the problem as a GBSS with a Laplacian mixing matrix. Since the topology recovery stage of the ML-BEST is shown to be a NP-hard optimization problem, we propose two low-complexity algorithms for the implementation of the topology recovery stage of the ML-BEST estimator: 
1) a two-phase topology recovery algorithm, which finds the relaxed positive semidefinite mixing matrix solution and then finds the closest Laplacian matrix to this solution by using convex optimization; 2) an augmented Lagrangian topology recovery algorithm, which is based on classical cICA approaches.  These methods rely only on the SOS of the state signals and, in contrast to classical BSS techniques, enable the separation of Gaussian sources. We present some identifiability conditions for this GBSS problem, complexity analysis of the proposed ML-BEST methods, and the associated CRB of the demixing parameters. 
 Numerical simulations show that the proposed ML-BEST methods succeed in reconstructing the topology and estimating the states, 
and that the topology estimators achieve the CRB asymptotically.
The augmented Lagrangian  ML-BEST may be preferable for large networks, 
since the two-phase ML-BEST is a computationally heavy algorithm, as described in Subsection \ref{sec_complex}.
Additionally,  the state estimators converge to the oracle state estimator, which assumes perfect knowledge of the topology. 

State estimation is the backbone of power system monitoring and processing. The presented results indicate that even if the topology recovery is not perfect, the MSE of the state estimation is close to the MSE of the oracle performance. Thus, the proposed ML-BEST methods can be applied for practical power system operations without assuming knowledge of the topology.
In future work, the proposed methods will be extended to address complex
 random states, by incorporating concepts from complex BSS \cite{Adali_Schreier_Scharf2011} and the proposed GBSS approach. 
For the sparsity pattern,  more general thresholding functions should be investigated, as well as theoretical recovery guarantees.

\end{document}